# Semi-Distributed Demand Response Solutions for Smart Homes

Rim Kaddah, Daniel Kofman, Fabien Mathieu and Michal Pióro[1]

**Abstract** The Internet of Things (IoT) paradigm brings an opportunity for advanced Demand Response (DR) solutions. It enables visibility and control on the various appliances that may consume, store or generate energy within a home. It has been shown that a centralized control on the appliances of a set of households leads to efficient DR mechanisms; unfortunately, such solutions raise privacy and scalability issues. In this chapter we propose an approach that deals with these issues. Specifically, we introduce a scalable two-levels control system where a centralized controller allocates power to each house on one side and, each household implements a DR local solution on the other side. A limited feedback to the centralized controller allows to enhance the performance with little impact on privacy. The solution is proposed for the general framework of capacity markets.

## 1 Introduction

The growing deployment of intermittent renewable energy sources at different scales (from bulk to micro generation) advocates for the design of advanced Demand Response (DR) solutions to maintain the stability of the power grid and to optimize the usage of resources.
DR takes advantage of demand flexibility, but its performance depends on the granularity of visibility and demand control. The Internet of Things (IoT) paradigm enables implementing DR at the finest granularity (individual appliances), and deploying IoT-based solutions becomes feasible, both from the technological and economical points of view.

[1] R. Kaddah (✉), D. Kofman
Telecom Paristech, 23 Avenue d'Italie, Paris, France
e-mail: {rim.kaddah, daniel.kofman}@telecom-paristech.fr
F. Mathieu
Nokia Bell Labs, route de Villejust, Nozay, France
e-mail: fabien.mathieu@nokia.com
M. Pióro
Institute of Telecommunications, Warsaw University of Technology, Poland
Department of Electrical and Information Technology, Lund University, Sweden
e-mail: michal.pioro@eit.lth.se

The introduction of capacity markets in several countries has provided incentives for: the flexibility end users could provide through DR mechanisms; the deployment of flexible generators (for which the energy cost is higher than the average).

In this chapter, we focus on DR solutions for keeping power consumption below a certain known capacity limit for a well-defined period. A possible application is for utility companies, which are interested in limiting the cost of the capacity certificates they have to acquire in the capacity market (for securing supply). Such a cost reduction is facilitated by keeping power consumption below known thresholds.

In [1], the authors propose and analyze several IoT-based DR mechanisms. They show that fine-grained visibility and control on a set of households at an aggregation point enables to maximize user's perceived utility. However, this approach may cause scalability as well as privacy problems. On the other hand, they consider two levels control systems where a central controller allocates available capacity to households based on some static information (e.g., type of contract). Then, local controllers leverage IoT benefits for local optimization, without any feedback to the central controller. The drawback of such approach is that it may reduce the total utility perceived by the users.

Our main contribution is a proposition and evaluation of an intermediate approach, based on two-level systems with partial feedback from the local controllers to the central entity, where the feedback sent has little impact on privacy. The proposed solution enforces fairness by considering two levels of utility for each appliance (i.e., vital and comfort). We compare the performance of the proposed scheme with the two cases studied in [1] (fully centralized solution and two level system with no feedback). The results are analyzed for homogeneous and heterogeneous scenarios. We show that for both cases, the proposed algorithm outperforms the scheme with no feedback. Moreover, it runs in a limited number of feedback iterations, which ensures good scalability and limited requirements in terms of communication.

The chapter is organized as follows: Section 2 presents the related work. The system model and allocation schemes are introduced in Section 3 and 4, respectively. Section 5 studies the performance of the proposed control scheme and compares it with two benchmark control approaches through a numerical analysis of the model. Conclusions and future work are presented in Section 6.

## 2 Related work

The idea of using vertically distributed control schemes with no or limited feedback is quite natural. Indeed, the literature contains a significant amount of pro-posals based on hierarchical control schemes (with feedback) in the context of limiting consumption capacity to a certain desired value (e.g., [2, 3, 4, 5]).

However, these proposals usually address the problem by taking the dual of system capacity constraint. These dual variables can be seen as prices for time slots (e.g., [2, 3, 4]). Thus, these schemes are usually presented as DR schemes based on pricing. In contrast, our approach is based on direct control of the appliances.

The proposals in [3, 4, 5] are examples of schemes that are designed for residential consumers and that can take into account flexibility of generic appliances.

Authors in [3] propose a customer reward scheme that encourages users to accept direct control of loads. They propose a time-greedy algorithm (maximizes utility slot by slot) based on the utility that each appliance declares for each slot. As discussed in the previous chapter, instantaneous value of an appliance has to be carefully evaluated to capture the real benefit from using this appliance (e.g., heating system). Authors in [4] propose a dynamic pricing scheme based on a distributed algorithm to compute optimal prices and demand schedules. Closer to our proposal is the direct control scheme presented in [5] which is very similar to [4] if prices are interpreted as control signals. The authors in [5] propose to solve a problem similar to ours, but in their approach, intermediate solutions can violate the constraints, so that convergence of the algorithm is required (like all other schemes based on dual decomposition) to produce a feasible allocation. The authors do not discuss scalability and communication requirements in terms of the number of iterations required. They also assume concave utility functions. Moreover, the proposed scheme still requires disclosure of extensive information to the central entity (i.e., home consumption profile), so the approach is not adapted to reduce privacy issues.

In the present work, we target to better deal with privacy while guaranteeing the fulfillment of capacity constraints even during intermediate computation. We achieve that by building on the DR framework introduced in [1].

## 3 System model

We consider an aggregator in charge of allocating power to a set of $H$ households under a total capacity constraint $C(t)$. $t$ represents a time slot. We suppose that during a defined time period (measured in slots), in absence of control, predicted demand would exceed available capacity. We call this period a DR period. We denote by $DE_a$ and $DE_h$ the functional groups in charge of decision taking (Decision Entities) at the aggregator side and at the user $h$ side (one per home), respectively. $DE_a$ is in charge of allocating power to each household ($C_{ht}$), under the total power constraint. For each house $h$, $DE_h$ has two main roles: collecting information on variables monitored at user premises (state of appliances, local temperature, etc.); enforcing control decisions received from $DE_a$ (e.g. by controlling the appliances). More details will be given in Section 4 when introducing the considered allocation schemes.

A utility function is defined for each controlled appliance to express the impact of its operation on user's satisfaction. We assume electrical appliances are classified among $A$ classes. Appliances of the same class have similar usage purposes (e.g., heating) but may have different operation constraints. Appliance of class $a$ at home $h$ operates within a given power range $[P_m^a(h), P_M^a(h)]$.

Following [1], a specific utility function is modeled for each class of appliances based on usage patterns, criticality, users' preferences and exogenous variables (e.g. external temperature). The utility of an appliance is expressed as a function of its consumption or of some monitored variables (see Section 5 for an example). In the present work, we introduce two levels of utility per appliance, vital and comfort. The first one expresses high priority targets of high impact on users' wellbeing and the second one expresses less essential preferences.

For notation, we write utilities as vital/comfort pairs: $U_{ht}^a = (U_{v_{ht}}^a, U_{c_{ht}}^a)$ denotes utility of appliance $a$ at time $t$ for home $h$. Control decisions are based on the lexicographical order comparison of utility values: higher vital value is always preferred regardless of the comfort value. Formally, for two utilities $U_{ht}^a$ and $U'^a_{ht}$, we say $U_{ht}^a > U'^a_{ht}$ iff $U_{v_{ht}}^a > U'^a_{v_{ht}}$ or ($U_{v_{ht}}^a = U'^a_{v_{ht}}$ and $U_{c_{ht}}^a > U'^a_{c_{ht}}$).
Utilities can be summed using element-wise addition.

**Table 1: Table of notation.**

| | |
|---|---|
| System Parameters and Exogenous Variables | |
| $H$ | Number of homes |
| $A$ | Number of appliance classes |
| $P_m^a(h)$ / $P_M^a(h)$ | Minimal / Maximal power required by appliance $a$ in home $h$ |
| $C(t)$ | Available power capacity at time slot $t$ |
| $L(h)$ | Subscribed power for home $h$ |
| $t_M$ | DR period duration in time slots |
| $T_m(h)$ / $T_M(h)$ | Minimal / Maximal acceptable indoor temperature for home $h$ |
| $T_0(h)$ / $T_P(h)$ | Initial / Preferred indoor temperature for home $h$ |
| $T_e(t)$ | Exterior temperature at time $t$ |
| $F(h), G(h)$ | Coefficients for temperature dynamics in home $h$ |
| Control Variables and Controlled Variables | |
| $U_{ht}^a = (U_{v_{ht}}^a, U_{c_{ht}}^a)$ | Utility (vital, comfort) of appliance $a$ in home $h$ at time $t$ |
| $X_{ht}^a$ | Power consumed by appliance $a$ in home $h$ at time $t$ |
| $x_{ht}^a$ | Activity indicator of appliance $a$ in home $h$ at time $t$ (0 or 1) |
| $T_{ht}$ | Temperature of home $h$ at time $t$ |
| $C_{ht}$ | Capacity limit allocated for home $h$ at time $t$ |
| $g_{ht}$ | Greedient of the utility function at point $C_{ht}$ for home $h$ at time $t$ |

The maximal values of utilities depend on the home, type of appliance and time: they represent how the importance of appliances is modulated depending on the preferences and service agreement of the users. We assume that each house has a subscribed power limit $L(h)$ sufficient to achieve a maximal utility.

The optimization problem considered in this chapter consists in maximizing the total utility (using the lexicographic total order) of users under system constraints. Fairness is introduced through the vital/comfort separation: no comfort power is allocated to any house if some vital need can be covered instead. We do not directly focus on revenues but expect that reaching maximal users' utility leads to maximal gains for all involved players. Utility companies can provide better services for a given total allocated power, which should translate into higher revenues, or reduce the expenses in the capacity market for a given level of service, which should reduce total costs. End users can save money due to attractive prices they get for participating to the service and adjusting energy consumption to their predefined policies. Notation is summarized in Table 1.

## 4 Allocation schemes

We present here two reference schemes that will be used for benchmarking purposes, along with our proposed solution.

### *4.1 Benchmark schemes*

The two following schemes were proposed in [1].

#### 4.1.1 Global Maximum Utility

The centralized global optimization is formulated by Equation (1).

$$\max_{X_{ht}^a, x_{ht}^a} \sum_{t=1}^{t_M} \sum_{h=1}^{H} \sum_{a=1}^{A} U_{ht}^a \qquad (1a)$$

$$s.t.$$

$$\sum_{h=1}^{H} \sum_{a=1}^{A} X_{ht}^a \le C(t), \forall t \qquad (1b)$$

$$P_m^a(h) x_{ht}^a \le X_{ht}^a \le P_M^a(h) x_{ht}^a, \ \forall\, t, \ \forall\, h, \ \forall\, a \qquad (1c)$$

$$x_{ht}^a \in \{0, 1\}, \ \forall\, t, \ \forall\, h, \ \forall\, a \qquad (1d)$$

Equation (1) can be solved if all information about appliances and their utility functions are transmitted by the home repartitors $DE_h$ to the aggregator $DE_a$, which can then compute an optimal global solution and notify the repartitors accordingly.

Decision variables in this case are variables $x_{ht}^a$ and $X_{ht}^a$. Binary variables $x_{ht}^a$ correspond to turning ON (i.e., $x_{ht}^a = 1$) or OFF (i.e., $x_{ht}^a = 0$) appliance a at home h on time slot t. If appliance is turned ON, power allocation $X_{ht}^a$ can take values between a minimum value $P_m^a(h)$ and a maximum value $P_M^a(h)$ (see equation (1c)).

While being optimal with respect to the utilities (by design), this allocation, called $GM$, has two major drawbacks. First, it requires computing the solution of a complex problem, which may raise scalability issues. Second, information harvesting may cause privacy issues that can affect the acceptance of the control scheme by users. Thus, it may be preferable to store information locally at homes with local intelligence. This leads to the following scheme.

**4.1.2 Local Maximum Utility**

This control scheme, denoted $LM$, considers only one-way communication from $DE_a$ to $DE_h$ (no feedback from $DE_h$ to $DE_a$). Decisions are made at both levels. First, $DE_a$ allocates power to homes proportionally to their subscribed power, so the power allocated to home $h$ is $C_{ht} = \frac{L(h)}{\sum_i L(i)} C(t)$.

Then, at each home $h$, $DE_h$ decides the corresponding allocation per appliance by solving the restriction of (1) to $h$, using $C_{ht}$ instead of $C(t)$.

By design, $LM$ is scalable (only local problems are solved) and private information disclosure is kept to a minimum. The drawback is that the corresponding allocation may be far from optimal [1].

## *4.2 Greedient approach*

We now propose a two-way scheme that aims at achieving a trade-off between performance, scalability and privacy.

To reach privacy and scalability goals with limited feedback, we propose a simple primal decomposition of the global $GM$ problem into a master problem, described in Equation (2), and subproblems, described in Equation (3).

**Master problem**

$$\max \sum_{h=1}^{H} U_h \quad (2a)$$
$$\sum_{h=1}^{H} C_{ht} = C(t), \forall t \quad (2b)$$
$$C_{ht} \geq 0, \forall h, \forall t \quad (2c)$$

**Subproblems**
For each home $h$, the following MILP is solved:

$$U_h = \max \sum_{t=1}^{t_M} \sum_{a=1}^{A} U_{ht}^a \quad (3a)$$
$$\sum_{a=1}^{A} X_{ht}^a \leq C_{ht}, \forall t \quad (3b)$$

If the $C_{ht}$ are known, the subproblems (3) can be solved like in the $LM$ scheme. The main issue is the master problem (2): how to shape an optimal per-home allocation while keeping the full characteristics of appliances private?

To treat this problem, we propose a new heuristic called the Sub-Greedient method ($SG$). This heuristic is inspired by the Sub-Gradient method [6], but is adapted to take into account the specificities of our model. In particular, we introduce the notion of *Greedient*, inspired by the gradient method and the metric used to sort items in the knapsack greedy approximation algorithm[2]. Greedients will be used instead of more traditional (sub)-gradient approaches to estimate the utility meso-slope of a given house.

We briefly describe the main steps of $SG$:

- $SG$ needs to be bootstrapped with an initial power allocation.
- $DE_a$ transmits to each home $DE_h$ the current allocation proposal $C_{ht} \forall\, t$. $DE_h$ then solves the corresponding subproblem (3). It sends back the total utility $U_h$ feasible, along with the Greedient associated to the current solution.
- Using the values reported by homes, $DE_a$ then tries to propose a better solution.
- The process iterates for up to $K_{MAX}$ iterations, and return the best solution found.

We now give the additional details necessary to have a full view of the solution.

**4.2.1 Initial allocation**

Following [1], we use a round-robin strategy for the first allocation (before the first feedback): we allocate to some houses up to their power limit until the available capacity $C(t)$ is reached; we cycle with time the houses that are powered. The interest for $SG$ of such an initial allocation (e.g. compared $LM$) is that it breaks possible symmetries between homes and gives an initial diversity that will help finding good Greedients.

**4.2.1 Greedient**

We define the greedient $g_{ht}$ as the best possible ratio between utility and capacity improvements of home $h$ at time $t$. Formally, if $U'_h(\Delta C_t)$ represents the best feasible utility for home $h$ if its current allocation is increased by $\Delta C_t$ at time $t$, we have

$$g_{ht} := \max_{\Delta C_t > 0} \frac{U'_h(\Delta C_t) - U_h}{\Delta C_t}.$$

To compute $g_{ht}$, we define the greedient $g_{ht}^a$ of an appliance $a$ as follows: for a given allocation $C_{ht}$, $C_{ht}^0 \geq 0$ represent the capacity unused by house $h$ at time $t$ in the optimal allocation. $U_h^{a'}(\Delta C_t)$ represents the maximum utility for appliance $a$ if

---

[2] The term *discrete gradient* could be used instead of this neologism. However, the greedient, which will be formally defined below, differs from the usual definition of a discrete gradient [7].

an additional capacity of up to $\Delta C_t$ is added its current consumption. Then we have

$$g_{ht}^a := \max_{\Delta C_t > 0} \frac{U_h^{a'}(C_{ht}^0 + \Delta C_t) - U_h^a}{\Delta C_t}.$$

The greedient of a home is the greedient of its best appliance : $g_{ht} = \max_a g_{ht}^a$.

Note that if we suppose that the utility functions have a diminishing return property, which is the case for our numerical analysis, the greedient of an appliance is equivalent to the gradient of the utility function when $C_{ht}^0 = 0$ and continuous variation of power is allowed: for these situations, the best efficiency is observed for $\Delta C_t \to 0$. The only difference (under diminishing return assumption) is when allowed allocations are discrete: the greedient will consider to the next allowed value while the gradient will report 0.

**Remark** The improvement advertised by the greedient is only valid for a specific capacity increase, which is not disclosed to $DE_a$ to prevent the central entity to infer the characteristics of users based on their inputs. As a result, the greedient hints at the potential interest of investing additional capacity to a given home, but it is not reliable. This is the price we choose to pay to limit privacy issues.

### 4.2.3 Finding better solutions

To update the current solution at the k-th iteration, $DE_a$ does the following:

- It first computes values $\alpha_k g_{ht} \ \forall \ h \ \forall \ t$. These values represent potential increase of $C_{ht}$. The values of $\alpha_k$, called the *step size*, are discussed below.
- It then adjusts the new values of $C_{ht}$. based on these values, while staying positive and fitting the capacity constraints.

For the adjustment phase, it is important to deal with cases where allocation update $\alpha_k g_{ht}$ is larger than available capacity $C(t)$ or even maximum subscribed power $L(h)$ of home $h$, so we first cap $\alpha_k g_{ht}$ at the minimum between power limit of the smallest home ($L_m := min_h L(h)$)[3] and system capacity $C(t)$. We therefore define $\beta_{kht} = min(\alpha_k g_{ht}, Lm, C(t))$.

Then for each $t$, we remove some positive common value $\lambda_t$ to the $C_{ht}$ to keep the sum of the allocations equal to the total capacity $C(t)$. To avoid houses with low $C_{ht}$ to be badly impacted (in particular to avoid negative allocations that will be impossible to enforce), a subset $I_t$ of the houses will be "protected" so that their values cannot decrease. In details, we do the following, starting with $I_t = \emptyset$:

- We compute $\lambda_t$ such that the values

---

[3] We chose the capacity of the smallest home instead of the capacity of the current home to avoid a *masking* effect where the demands of larger homes cloud the demands of smaller homes.

$$C'_{ht} = \begin{cases} C_{ht} + max\{\beta_{kht} - \lambda_t, 0\} \text{ if } h \in I_t, \\ C_{ht} + \beta_{kht} - \lambda_t \text{ otherwise,} \end{cases}$$

sum to $C(t)$. See [8, 9] for more details.

- We protect (e.g. add to $I_t$) all houses that get a negative value $C'_{ht}$.
- We iterate the steps above until all $C'_{ht}$ from eq. (4) are positive. $DE_a$ then proposes $C'_{ht}$ as a new solution to investigate.

**Remark** The solution described here applies to a 2-level hierarchy ($DE_a, DE_h$), but it can be generalized to $M$ levels to take into account different aggregation points on a hierarchical distribution network: considering an aggregation point $m$ at a certain level, the greedient for $m$ is the maximal greedient of its children. The adjustment phase can take into account capacity constraints of $m$, such as static power limits at each level of the hierarchical distribution network.

Also note that the proposed scheme does not require all houses to communicate simultaneously: it can run asynchronously. In fact, as soon as at least two homes respond, a local reallocation can be made: we just need to restrict the problem to the corresponding subset of homes, using their current cumulated allocation as capacity limit.

### 4.2.4 Choosing the step size

The step size $\alpha_k$ for each iteration $k$ is a crucial parameter to speed up resolution. Intuitively, large values of $\alpha_k$ make the allocation update (dictated by $\alpha_k g_{ht}$) useful for high consumption appliances, while lower values are more adapted to low consumption appliances.

Among the step size sequences proposed for subgradient methods, we consider for our performance analysis the two following ones (see [6]):

A diminishing non-summable step size rule   of the form $\alpha_k = \frac{a_1}{\sqrt{k}}$.

A constant step length rule   of the form $\alpha_k = \frac{a_2}{\|g_{ht}\|_2}$, where $\|g_{ht}\|_2$ is the euclidean norm of the vector of all greedients.

The value of parameter $a_1$ (resp. $a_2$) is currently manually adjusted to provide the best result, but we believe that an automatic estimation of the best value given the static parameters of a given use case is a promising lead for future work.

# 5 Numerical analysis

We now evaluate the performance of our proposed solution for a specific use case.

## *5.1 Parameters and settings*

We consider three typical types of appliances ($A = 3$): lighting ($a = 1$), heating (index $a = 2$) and washing machines (index $a = 3$). For these appliances, user's perceived utility respectively depends on: instantaneous power consumption; exogenous variables (temperature); the completion of a program. Utility functions for these appliances have a vital and a comfort component. For lighting, vital light utility is fully obtained as soon as the minimal light power $P_m^1(h)$ is reached, while comfort utility linearly grows from $P_m^1(h)$ to $P_M^1(h)$ (see Figure 1). For heating, vital utility linearly grows until the minimum tolerable temperature $T_m(h) := 15°C$ is reached, while comfort utility linearly grows from $T_m(h)$ to the preferred temperature $T_P(h) := 22°C$ (see Figure 2). For washing machines, an operation of duration $D(h)$ needs to be scheduled between an earliest start time $t_s(h)$ and a deadline $t_d(h)$. Once started, an operation cannot be interrupted. Vital utility function is maximal whenever the operation is successfully scheduled, while comfort utility depends on the execution time, e.g. the sooner the better for this use case (See Figure 3).

To study the performance of the control schemes for several values of capacity, we choose the following system parameters:

- The size of the system is $H = 100$ houses.
- We select a slot duration of 5 minutes.
- The DR period is set to $t_M = 100$ slots ($\cong 8$ hours).
- We suppose a constant external temperature $T_e(t) = 10°C \: \forall \: t$ and an initial temperature $T_0(h) = 22°C \: \forall \: h$.
- We suppose the same maximal utility values for all appliances, homes and time, arbitrary set to 1.
- Temperature in homes evolves according to a simplified conductance/capacity model that leads to the following dynamics:
  $T_{ht} = T_{h(t-1)} + F(h)X_{ht}^2 + G(h)(T_e(t) - T_{h(t-1)})$.
- Two types of houses are considered (See Tables 2-5). Compared to class 1, class 2 has a better energetic performance (less light power required, better insulation and more efficient washing machine), resulting in a lower power limit $L(h)$).

We suppose that the total available power is constant over the DR period, $C(t) = C$. We analyze the model for different values of $C$, ranging from low (only one type of appliances can be used) to full capacity (all appliances can be used).

Table 2: Lighting parameters.

| Type | $P_m^1(h)$ | $P_M^1(h)$ |
|---|---|---|
| 1 | 50 | 1000 |
| 2 | 50 | 500 |

Table 3: Heating parameters.

| Type | $P_m^2(h)$ | $P_M^2(h)$ | F(h) | G(h) |
|---|---|---|---|---|
| 1 | 1000 | 4000 | 0.0017 | 0.075 |
| 2 | 1000 | 2000 | 0.0008 | 0.0365 |

Table 4: Washing machine parameters.

| Type | $P_m^3(h) = P_M^3(h)$ | D(h) | $t_s(h)$ | $t_d(h)$ |
|---|---|---|---|---|
| 1 | 600 | 8 | 1 | 100 |
| 2 | 400 | 6 | 1 | 100 |

Table 5: Houses parameters.

| Class | Lighting type | Heating type | Washing machine type | L(h) |
|---|---|---|---|---|
| 1 | 1 | 1 | 1 | 5600 |
| 2 | 2 | 2 | 2 | 2900 |

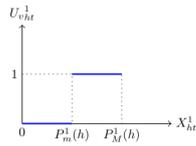
(a) Vital utility

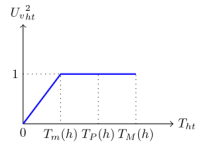
(a) Vital utility

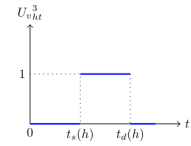
(a) Vital utility

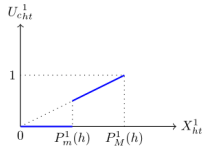
(b) Comfort utility

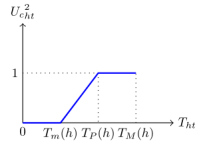
(b) Comfort utility

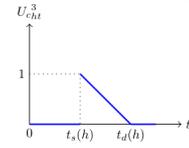
(b) Comfort utility

**Fig. 1: Utility of light power.**   **Fig. 2: Utility of $T_{ht}$.**   **Fig. 3: Utility of a washing machine.**

While this model is simple (three types of appliances, constant values), we believe that the knowledge required to compute good solutions is sufficient to capture the trade-off between the efficiency of an allocation and the privacy of the users.

For the Sub-Greedient problem, we fix the maximum number of iterations to $K_{MAX} = 100$. Two variants are considered (cf Section 4.2.4): $SG-1$ uses a diminishing step ($a_1 = 1200000$) and $SG-2$ uses a constant step length ($a_2 = 6000$). Parameters $a_1$ and $a_2$ were manually tuned.

The numerical analysis of the various presented mixed integer linear problems has been carried out using IBM ILOG CPLEX ([10]).

In the following, we discuss two cases: homogeneous and heterogeneous. For the homogeneous case, all houses belong to class 1 and for the heterogeneous one, we suppose 50 houses of class 1 and 50 houses of class 2.

## 5.2 Results on the homogeneous case

The main results on the homogeneous case are presented in Figure 4. It displays the relative utility per home over the DR period as a function of the available capacity $C$, for the four supposed schemes: $GM$, $LM$, $SG-1$ and $SG-2$.

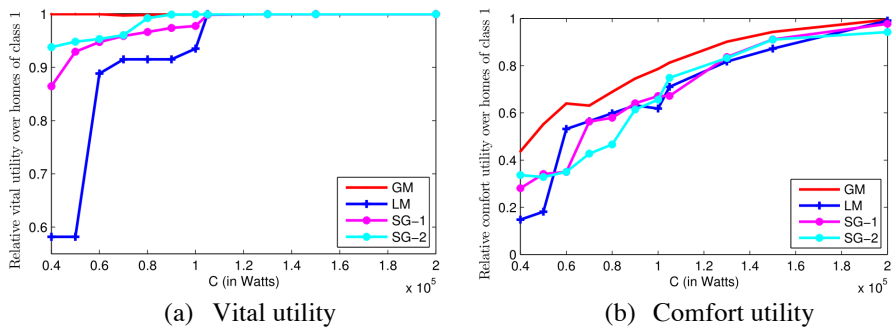

(a) Vital utility      (b) Comfort utility

**Fig. 4: Relative utility as a function of the available capacity (homogeneous case, class 1).**

The maximal feasible utility (vital and comfort) is normalized to 1 which is reached when all appliances from all homes of a given class reach their maximal utility. Another value of interest for vital utility is 0.58, which corresponds to situations where all houses are able to achieve vital light ($P_m^1 = 50$ W) but none has the power necessary for heating ($P_m^2 = 1000$ W) so there is no control of temperature, nor washing machines are scheduled. When washing machine are scheduled in addition to lights (without heating), vital utility reaches 0.92.

$GM$, the optimal solution, achieves maximal vital utility even for very low capacities (down to $4\times 10^4$), thanks to its ability of finding a working rolling allocation that allows all houses to use heat for a sufficient part of the period while scheduling the other appliances. Based on $GM$ results, we can measure the gap between optimal allocation and allocations obtained with $LM$, $SG-1$ and $SG-2$.

Using a static allocation, $LM$ struggles for rising the vital utility above the 0.58 and 0.92 thresholds. It can schedule washing machines when $C \geq 6\times 10^4$ ($P_m^3 = 600$ W per home). It can only start to use heat for $C = 10^5$ (1000 W per house). Maximal vital utility is reached for $C = 105\times 10^3$ (1050 W per house) and maximal utility (vital and comfort) necessarily requires $C = 2\times 10^5$ (2000 W per house). However, $LM$ achieves descent performance results for high enough available capacity values.

Our proposal, $SG-1$ and $SG-2$, stands in-between the two opposite schemes $GM$ and $LM$. Indeed, for very low capacity values ($C < 2\times 10^4$ W), $SG-1$ and $SG-2$ perform slightly better than $LM$. Then, for capacity values up to $C = 6\times 10^4$ W, $SG-1$ and $SG-2$ significantly over-perform $LM$. In particular, the schemes manage to activate most washing machines starting from $C = 4\times 10^4$ W (400W per home on average, to compare with the 600W required to operate a washing machine). It is able to improve the vital utility of houses for values below $C = 10^5$, even if it fails to perform as good as $GM$. With respect to the comfort utility, it performs on par with $LM$ even in situation where it devotes resources on heating (for vital utility) while $LM$ does not.

As for the number of iterations required to reach the best solution, $SG-1$ takes around 12 iterations on average and $SG-2$ takes 21 iterations. The slowest convergence is observed for capacity $C = 2\times 10^5$ W (maximal considered $C$) where $SG-1$ takes 95 iterations and $SG-2$ takes 98 iterations.

## *5.3 Results on the heterogeneous case*

Figure 5 illustrates the main results for the heterogeneous case for homes of classes 1 and 2. The optimal solution given by $GM$ shows that, for vital utility, the results are pretty much similar for both classes to the homogeneous case, with maximal value obtained even for low capacities (down to $4\times 10^4$). For the comfort utility, however, $GM$ leads to better values for class 2 compared to class 1. This is due to the fact that class 2 houses have better energetic performance, so once vital utility is ensured for all, it is more efficient to allocate energy to homes of class 2.

The same reason explains the poor performance of $LM$. Let us remember that the static allocation is proportional to the maximum power $L(h)$ of homes. So for a given capacity, class 1 homes get more power than class 2 ones. As a result, while performance of class 1 is satisfactory, performance of class 2 is terrible despite the better energy performance of class 2 homes. In particular, the capacity required for class 2 houses to achieve maximal vital utility is very high: $C = 1.7 \times 10^5$, which corresponds to 1700 W per house (regardless the class).

For lower capacity values, performance depend on the possibility of scheduling washing machines and heating. A global capacity $C = 5 \times 10^4$ will only allow homes of class 1 to schedule their washing machines. Actually, for this capacity value, homes of class 1 will get a power limit of 659 W whereas homes of class 2 will only get 341 W (insufficient for turning on a washing machine). For capacity values above $C = 7 \times 10^4$ (corresponding to slightly more than 450 W for each home of class 2), performance obtained corresponds to washing machines being scheduled and minimum lighting requirements being fulfilled for both classes while only homes of class 1 have their vital heating requirement.

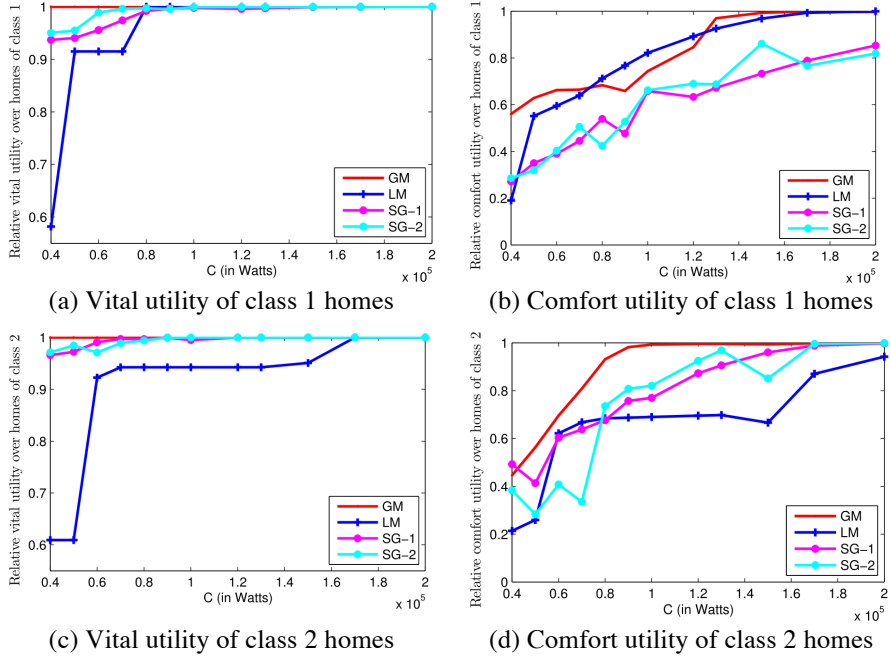

(a) Vital utility of class 1 homes  
(b) Comfort utility of class 1 homes  
(c) Vital utility of class 2 homes  
(d) Comfort utility of class 2 homes

**Fig. 5: Relative utility as a function of the available capacity (heterogeneous case, classes 1 & 2).**

As for $SG-1$ and $SG-2$, we observe that compared to the homogeneous case, the performance of our solution $SG$ is now closer to $GM$ than to $LM$. Indeed, $SG$ is capable of providing near maximal vital utility for $C = 0.6 \times 10^5$ W.

As for the number of iterations corresponding to the last solution improvement, $SG-1$ takes up to 3 iterations and $SG-2$ takes 14 iterations on average.

## *5.4 Discussion*

The results presented previously can be seen as intuitive. However, they show some interesting tradeoffs that need to be considered when proposing a DR solution. As suggested by the results, a fine grained control may not be always needed depending on the system's available capacity: its value is the highest for very low capacities. As a matter of fact, a solution based on static information can have high performance thanks to the deployment of a fine grained solution in smart-homes that can manage to efficiently schedule appliances based on user's needs when capacity is high enough.

However, producing an efficient solution based on static information is challenging especially when considerations like heterogeneity in users' needs and time-dependent constraints for appliances (e.g, minimum duration of operation) are supposed. In addition, one may imagine that available capacity will vary in time which also increases complexity of finding such a solution.

To address the lack of visibility while preserving privacy, a solution that uses limited information and is able to update allocations based on actual needs is needed. Actually, if high performance can be delivered by such a solution, a centralized approach will not be required.

The hierarchical solution proposed in this chapter is a promising one that is capable of addressing this need. It also seems to deal well with appliances introducing time dependence between time slots even if it is not a built in feature. It is capable of rendering a performant solution is a reasonable amount of iterations.

## 6 Conclusions

We propose an IoT-based demand response approach, named *Sub-Greedient*, that relies on a 2 level control scheme. Intelligence (decision taking) is split between a centralized component and a set of local controllers (one per home). The proposed control approach enables reaching good performance in terms of the utility perceived by the users while keeping privacy and providing scalability. Moreover, priority is provided for critical needs, which introduces some degree of fairness among households.

We show that the approach outperforms schemes where the central controller takes decisions based solely on the available total capacity and on static (contract-based) information about the households. Results for the considered use cases show that the proposed scheme requires a limited number of iterations to render effective solutions. Moreover, the proposed solution is robust as the algorithm stays inside the set of feasible allocations and can tolerate lost or delayed information.

Future work will encompass a study on the power allocation algorithms for the *Sub-Greedient* scheme considering the effect of communication impairments on the global performance and on fairness. We will analyze the cost savings under realistic cost models, looking for solutions that will target minimizing the total expenses a provider will incur in the capacity market while keeping a predefined level of service.

## Acknowledgement

The work presented in this chapter has been partially carried out at LINCS (www.lincs.fr)

## References

1. Kaddah R, Kofman D, Pióro M. Advanced demand response solutions based on fine-grained load control. In *IEEE International Workshop on Intelligent Energy Systems*, pages 38–45, San Diego (USA), October 2015.
2. Deng R, Lu R, Xiao G, Chen J. Fast Distributed Demand Response with Spatially-and Temporally-Coupled Constraints in Smart Grid. *IEEE Transactions on Industrial informatics,* 11(6):1597-1606, 2015.
3. Vivekananthan C, Mishra Y, Ledwich G, Li F. Demand Response for Residential Appliances via Customer Reward Scheme. *IEEE Transactions on Smart Grid*, 5(2):809–820, 2014.
4. Li N, Chen L, Low SH. Optimal demand response based on utility maximization in power networks. In *IEEE Power and Energy Society General Meeting*, pages 1-8, Detroit (USA), July 2011.
5. Shi W, Li N, Xie X, Chu C-C, Gadh R. Optimal residential demand response in distribution networks. *IEEE journal on selected areas in communications*, 32(7):1441-1450, 2014.
6. Boyd S, Xiao L, Mutapcic A. Subgradient methods. *lecture notes of EE392o,* Stanford University (USA), Autumn Quarter, 2003.
7. Bagirov AM, Karasözen B, Sezer M. Discrete gradient method: derivative-free method for nonsmooth optimization. *Journal of Optimization Theory and Applications*, 137(2):317–334, 2008.
8. Pióro M, Medhi D. *Routing, Flow and Capacity Design in Communication and Computer Networks*. Morgan Kaufmann Publishers, 2004.
9. Held M, Wolfe P, Crowder HP. Validation of subgradient optimization. *Mathematical programming,* 6(1):62–88, 1974.
10. IBM Ilog CPLEX optimizer. http://www-01.ibm.com/software/commerce/-optimization/cplex-optimizer/